\documentstyle[sprocl]{article}

\def\Journal#1#2#3#4{{#1} {\bf #2}, #3 (#4)}

\def\NCA{{\em Nuovo Cimento} A}
\def\NPB{{\em Nucl. Phys.} B}
\def\PLB{{\em Phys. Lett.}  B}

\def\PRD{{\em Phys. Rev.} D}

\def\JETPL{\em JETP Lett.}
\def\CQG{\em Class. Quantum Grav.}
\def\AP{{\em Ann. Phys.} (N.Y.)}
\def\IJMP{{\em Int. J. Mod. Phys.} A}

\def\be{\begin{equation}}
\def\ee{\end{equation}}
\def\bea{\begin{eqnarray}}
\def\eea{\end{eqnarray}}

\newcommand {\cA}{{\cal A}}
\newcommand {\cB}{{\cal B}}

\newcommand {\cL}{{\cal L}}
\newcommand {\cM}{{\cal M}}

\def\a{\alpha}

\def\d{\delta}

\def\g{\gamma}
\def\G{\Gamma}

\def\j{\psi}

\def\l{\lambda}

\def\o{\omega}

\def\z{\zeta}

\def\J{\Psi}

\begin{document}

\begin{flushright}
TSU-QFT-11/96 \\
hep-th/9608164\\
\end{flushright}
\vspace{10mm}

\title{N--EXTENDED LOCAL SUPERSYMMETRY OF MASSLESS PARTICLES
IN SPACES OF CONSTANT CURVATURE \footnote{Talk given at the
Second International Sakharov Conference on Physics, Lebedev Physical
Institute, Moscow, May 20--24, 1996.}}

\author{ S.M. KUZENKO }

\address{Department of Physics, Tomsk State University, Lenin Ave. 36,\\
Tomsk 634050, Russia}

\maketitle\abstracts{
We review the unified description of massless spinning particles, living
in spaces of constant curvature, in the framework of the pseudoclassical
approach with a gauged $N$-extended worldline supersymmetry and
a local $O(N)$ invariance.
}

In the pseudoclassical approach \cite{half}, the spin degrees of
freedom of point particles are realized by anticommuting variables
which turn into a set of generalized $\g$-matrices at the quantum level.
This approach is essentially supersymmetric, since the consistent
treatment of a particle with spin requires twice as many of local
worldline supersymmetries as the value of spin.

The mechanics action with a
gauged $N$-extended supersymmetry for a massless particle in Minkowski
space was suggested some years ago by Gershun and Tkach \cite{gt} and
investigated in detail by Howe {\it et al.} \cite{hppt}. In particular,
it was argued that worldline supersymmetry is compartible with
arbitrary gravitational background only for $N \leq 2$. This bound is
very natural because of the known problems with formulating
the higher-spin dynamics in curved space. Such problems do not
in general arise when the background geometry is
chosen to be maximally symmetric, although it was believed \cite{hppt} for
a time that Minkowski space is the only background compartible
with worldline supersymmetry for $N > 2$. In a recent paper \cite{ky} we have
extended the Gershun-Tkach (GT) model \cite{gt} to the cases of de Sitter
(dS) and anti-de Sitter (AdS) spaces. Our construction provides a
unified treatment of the dynamics of massless particles in spaces of
constant curvature and is based on a hidden conformal invariance.

Howe {\it et al} \cite{hppt} demonstrated that in $d=3+1$
dimensions the wave functions in the GT model satisfy the
conformally covariant equation
for a pure spin-$\frac{1}{2}N$ field strength
(helicities $\pm \frac{1}{2}N$) \cite{pr}.
That might apparently have implied  conformal invariance of
the model for all $d$ and $N$. This proposal has been proved by
Siegel \cite{s1} who found the ansatz to obtain the GT model from an
explicitly conformal ($O(d,2)$ invariant)
mechanics action in $d$ space and 2 time dimensions
(Siegel extended, to the higher-spin case,
the construction originally used by
Marnelius \cite{m} to represent the actions for  massless spin-0 and
spin-$\frac{1}{2}$
particles in a manifestly conformal form). It turns out  \cite{ky}
that the same $(d+2)$-dimensional action can be used to derive the
point particle models with $N$-extended worldline supersymmetry
in the dS and AdS spaces.

We consider the mechanics system in $d$ space and 2 time
dimensions with the action \cite{ky,s1}
$S = \int{\rm d}\tau \cL$ given by
\be
\cL = \frac{1}{2}\dot{Z}^\cA \dot{Z}_\cA + \frac{{\rm i}}{2}{\Gamma_i}^\cA
\dot{\Gamma}_{i\cA} - \frac{{\rm i}}{2}\varphi_{ij}{\Gamma_i}^\cA
\Gamma_{i\cA}\;.
\label{1}
\ee
Here  $\varphi_{ij}(\tau)$, $\varphi_{ij} = - \varphi_{ji}$,
are Lagrange
multipliers,
the bosonic $Z^{\cA}(\tau)$, $\cA = d+1,0,1,\ldots,d$,
and fermionic ${\Gamma_i}^{\cA}(\tau)$,
$i =1,\ldots,N$, dynamical variables are subject to the constraints
\bea
\eta_{\cA\cB}Z^{\cA} Z^{\cB} &=&0, \;\;    \qquad Z \neq 0  \label{2}      \\
\eta_{\cA\cB}Z^{\cA} {\Gamma_i}^\cB& =&0
\label{3}
\eea
with $\eta_{\cA\cB} = {\rm diag}(--+\ldots+)$.
Hence the variables $Z^\cA$ parametrize
the cone $Q$ in ${\bf R}^{d,2}$, whilst
${\Gamma_i}^\cA$ form $n$ tangent vectors to point $Z$ of the cone.

Along with the explicit global $O(d,2)$ invariance (conformal invariance),
the model possesses a rich gauge structure. The action remains unchanged
under worldline reparametrizations
and local $O(N)$ transformations \cite{ky,s1}.
Moreover, the action is invariant under local $N$-extended supersymmetry
transformations of rather unusual form \cite{ky}. These transformations
involve an $external$ $(d+2)$-vector $W^\cA$, chosen to satisfy
the only requirement
$(Z,W) = Z^\cA W_\cA  \neq 0$
for the worldline $\{Z^\cA (\tau), {\Gamma_i}^\cA (\tau),
\varphi_{ij} (\tau)\}$
in field, and read as follows
\bea
\delta {\Gamma_i}^\cA& = &Z^\cA \stackrel{\bullet}{\alpha}_i - \dot{Z}^\cA
\alpha_i + \frac{{\rm i}}{(Z,W)} {\Gamma_i}^\cB \Gamma_{j\cB} \alpha_j
W^\cA ,\nonumber\\
\delta Z^\cA& = &{\rm i}\alpha_i {\Gamma_i}^\cA, \qquad
\delta \varphi_{ij} = - \frac{{\rm i}}{(Z,W)} \alpha_{[i}
{\stackrel{\bullet}{\Gamma}_{j]}}\!\,^\cA W_\cA\;.
\label{4}
\eea
Here $\stackrel{\bullet}{\alpha}_i$ denotes an $O(N)$-covariant derivative,
$\stackrel{\bullet}{\alpha}_i = \dot{\alpha}_i - \varphi_{ij} \alpha_j$,
and similarly for ${\stackrel{\bullet}{\Gamma}_i}\!\,^\cA$. The origion of
the last term in $\delta \Gamma$ is to preserve the constraint (\ref{3}).

The expressions (\ref{4}) become $W$-independent only on the mass shell.
Off-shell, however, the supersymmetry transformations do not commute
with the conformal ones, in spite of the manifest $O(d,2)$ invariance of
$\cL$\,!
What is the physical origion of the presence of $W$-terms in (\ref{4})?
It turns out that the fixing of $W$ breaks the $O(d,2)$-invariance
and uniquely specifies some $d$-dimensional spacetime which is embedded into
the compact projective space $PQ$ related to
the cone (\ref{2}). $PQ$ is defined as the set of straight lines through
the origion of the cone. Associated to a non-zero $(d+2)$-vector $W$ is
the $d$-dimensional open submanifold $\cM_W$ in $PQ$
\be
\cM_W = \{\bar{Z}^\cA \in PQ, \;\;\; e^{-1} \equiv (Z,W)^2 > 0 \}
\label{5}
\ee
which can be parametrized by constrained $d+2$ projective variables
of the form
\be
\z^\cA = \frac{Z^\cA}{(Z,W)},\; \qquad \z^2 = 0,\; \qquad (Z,W) = 1.
\label{6}
\ee
Introducing on $\cM_W$ the metric
${\rm d}s^2 = {\rm d}\z^\cA{\rm d}\z_\cA = e{\rm d}Z^\cA {\rm d}Z_\cA$,
$\cM_W$ turns into a spacetime of constant curvature.
Three inequivalent choices for $W$:
$$
W^{\cA}_{(M)} = ( -\frac{1}{\sqrt{2}},0,\ldots,0,\frac{1}{\sqrt{2}}),
\;\;
W^{\cA}_{(AdS)} = (0,\ldots,0,\frac{1}{r}),
\;\;
W^{\cA}_{(dS)} = (\frac{1}{r},0,\ldots,0)
$$
leads to Minkowski, de Sitter (dS) and anti-de Sitter (AdS) spacetimes,
respectively; $(\pm 12r^{-2})$ being the curvature of the dS (AdS)
space.
The stability group of $W^\cA$ in $O(d,2)$ is seen to be
the symmetry group of the corresponding spacetime. With respect to the
symmetry group,
${\Gamma_i}^\cA$ is naturally decomposed as follows
\be
\l_i = e(\G_i,W), \; \qquad
{\J_i}^\cA = {\G_i}^\cA - (\G_i,W)\frac{Z^\cA}{(Z,W)}\;.
\label{7}
\ee
Eqs. (\ref{5}--\ref{7}) define the reduction of the conformal model (\ref{1})
to $d$ spacetime dimensions. The variables $e$ and $\l_i$
proves to enter the final Lagrangian as the einbein and
$N$-extended worldline gravitino respectively.

As an illustration, let us apply the reduction procedure described
to the case of the AdS
space. This space can be parametrized by $d+1$ constrained variables
$y^A \equiv \z^A$, where $A=d+1,0,1,\ldots,d-1$ (note $\z^d=r$). For
fermionic variables one gets
\be
\l_i = \frac{1}{r}e{\G_i}^d, \qquad
{\J_i}^A =  {\G_i}^A - \frac{1}{r}y^A {\G_i}^d, \qquad {\J_i}^d = 0.
\label{8}
\ee
The bosonic $y^A$ and fermionic ${\J_i}^A$ degrees of freedom are
constrained by
\be
y^Ay_A = -r^2, \;\; \qquad
y^A\J_{iA} = 0 \;.
\label{9}
\ee
Thus $\J_i$ present themselves $N$ tangent vectors to point $y$ of the AdS
hyperboloid. Now, the Lagrangian turns into
\be
\cL_{AdS} = \frac{1}{2e}(\dot{y}^A - {\rm i}\l_i{\J_i}^A)
(\dot{y}_A - {\rm i}\l_j \J_{jA})
 + \frac{{\rm i}}{2}{\J_i}^A(\dot{\J}_{iA} - f_{ij}\J_{jA}) \;.
\label{10}
\ee
where we have redefined
$\varphi_{ij} = f_{ij} + \frac{{\rm i}}{e}\l_i\l_j$.
The supersymmetry transformation (\ref{4}) takes the form
\bea
\delta y^A& = &{\rm i}\alpha_i {\J_i}^A, \qquad
\delta {\J_i}^A = - \frac{1}{e}\alpha_i(\dot{y}^A - {\rm i}\l_j
{\J_j}^A) -\frac{{\rm i}}{r^2}y^A{\J_i}^B\J_{jB}\a_j , \nonumber \\
\delta e& = &2{\rm i}\l_i\a_i, \qquad \d\l_i = \dot{\a_i} - f_{ij}\a_j,
\qquad
 \d f_{ij} = -\frac{{\rm i}}{r^2}\a_{[i}\J_{j]A}\dot{y}^A     \;.
\label{11}
\eea

It is of interest to reformulate the model in terms of internal
(unconstrained) coordinates $x^m$, $m=0,1,\ldots,d-1$, on the AdS space.
Then $\cL_{AdS}$ takes the form \cite{ky}
\bea
\cL_{AdS}& = &\frac{1}{2e}g_{mn}(\dot{x}^m - {\rm i}\l_i{\j_i}^a {e_a}^m)
(\dot{x}^n - {\rm i}\l_j{\j_j}^b {e_b}^n) \nonumber \\
{}& + & \frac{{\rm i}}{2}{\j_i}^a(\dot{\j}_{ia} - f_{ij}\j_{ja}
+ \dot{x}^m{\o_{ma}}^b\j_{ib}) \;.
\label{12}
\eea
Here $g_{mn}$ is the metric of the AdS space,
${e_m}^a$ and $\o_{mab}=-\o_{mba}$ its vielbein and torsion-free spin
connection, respectively; $a,b$ are tangent-space indices,
$a,b=0,1,\ldots,d-1$.
The unconstrained fermionic variables
${\j_i}^a$ carry a tangent-space vector index
and are defined by the rule $\j_{ia}
={e_a}^m \frac{\partial y^B}{\partial x^m} \J_{iB}$.
Remarkably, $\cL_{AdS}$ presents itself a minimal covariantization of the
flat-space Lagrangian  \cite{gt}. The supersymmetry transformations
inevitably involve, however, curvature-dependent terms \cite{ky}.

\section*{Acknowledgments}
I would like to thank my coworker J.V. Yarevskaya and gratefully
acknowledge fruitful discussions with N. Dragon, S.J. Gates and
O. Lechtenfeld.
This work was supported in part by the Alexander von Humboldt
Foundation and by the Russian Basic Research Foundation,
grant 96-02-16017.


\section*{References}
\begin{thebibliography}{99}
\bibitem{half}F.A. Berezin and M.A. Marinov, \Journal{\JETPL}{21}{543}{1975};
\Journal{\AP}{104}{336}{1977};
R. Casalbuoni, \Journal{\NCA}{33}{369}{1976};
A. Barducci, R. Casalbuoni and L. Lusanna,
\Journal{\NCA}{35}{377}{1976};
L. Brink, S. Deser, B. Zumino, P. Di Vecchia and P. Howe,
\Journal{\PLB}{64}{435}{1976};
L. Brink, P. Di Vecchia and P. Howe,
\Journal{\NPB}{118}{76}{1977}.

\bibitem{gt}V.D. Gershun and V.I. Tkach, \Journal{\JETPL}{29}{288}{1979}.

\bibitem{hppt}P. Howe, S. Penati, M. Pernicci and P. Townsend,
\Journal{\PLB}{215}{555}{1988}; \Journal{\CQG}{6}{1125}{1988}.

\bibitem{ky}S.M. Kuzenko and J.V. Yarevskaya, preprint ITP-UH-29/95,
hep-th/9512115; to appear in {\em Mod. Phys. Lett. A}.

\bibitem{pr}R. Penrose and W. Rindler, {\em Spinors and Spacetime}
(Cambridge Univ. Press, Cambridge, 1984).

\bibitem{s1}W. Siegel, \Journal{\IJMP}{3}{2713}{1988}.

\bibitem{m}R. Marnelius, \Journal{\PRD}{20}{2091}{1979}.


\end{thebibliography}
\end{document}